\PassOptionsToPackage{unicode}{hyperref}
\PassOptionsToPackage{hyphens}{url}
\documentclass[
  12pt,
  letterpaper,
]{article}
\usepackage{amsmath,amssymb}
\usepackage[]{ebgaramond}
\usepackage{iftex}
\ifPDFTeX
  \usepackage[T1]{fontenc}
  \usepackage[utf8]{inputenc}
  \usepackage{textcomp} 
\else 
  \usepackage{unicode-math}
  \defaultfontfeatures{Scale=MatchLowercase}
  \defaultfontfeatures[\rmfamily]{Ligatures=TeX,Scale=1}
\fi
\IfFileExists{upquote.sty}{\usepackage{upquote}}{}
\IfFileExists{microtype.sty}{
  \usepackage[]{microtype}
  \UseMicrotypeSet[protrusion]{basicmath} 
}{}
\makeatletter
\@ifundefined{KOMAClassName}{
  \IfFileExists{parskip.sty}{%
    \usepackage{parskip}
  }{
    \setlength{\parindent}{0pt}
    \setlength{\parskip}{6pt plus 2pt minus 1pt}}
}{
  \KOMAoptions{parskip=half}}
\makeatother
\usepackage{xcolor}
\usepackage{longtable,booktabs,array}
\usepackage{calc} 
\usepackage{etoolbox}
\makeatletter
\patchcmd\longtable{\par}{\if@noskipsec\mbox{}\fi\par}{}{}
\makeatother
\IfFileExists{footnotehyper.sty}{\usepackage{footnotehyper}}{\usepackage{footnote}}
\makesavenoteenv{longtable}
\usepackage{graphicx}
\makeatletter
\def\maxwidth{\ifdim\Gin@nat@width>\linewidth\linewidth\else\Gin@nat@width\fi}
\def\maxheight{\ifdim\Gin@nat@height>\textheight\textheight\else\Gin@nat@height\fi}
\makeatother
\setkeys{Gin}{width=\maxwidth,height=\maxheight,keepaspectratio}
\makeatletter
\def\fps@figure{htbp}
\makeatother
\newlength{\cslhangindent}
\newlength{\csllabelwidth}
\newlength{\cslentryspacingunit} 
\newenvironment{CSLReferences}[2] 
 {
  \setlength{\parindent}{0pt}
  \ifodd #1
  \let\oldpar\par
  \def\par{\hangindent=\cslhangindent\oldpar}
  \fi
  \setlength{\parskip}{#2\cslentryspacingunit}
 }%
 {}
\usepackage{calc}

\newcommand{\CSLLeftMargin}[1]{\parbox[t]{\csllabelwidth}{#1}}
\newcommand{\CSLRightInline}[1]{\parbox[t]{\linewidth - \csllabelwidth}{#1}\break}

\usepackage{siunitx}
\makeatletter
\@ifpackageloaded{subfig}{}{\usepackage{subfig}}
\@ifpackageloaded{caption}{}{\usepackage{caption}}
\AtBeginDocument{%

}
\AtBeginDocument{%

}
\newcounter{pandoccrossref@subfigures@footnote@counter}
\newenvironment{pandoccrossrefsubfigures}{%
\setcounter{pandoccrossref@subfigures@footnote@counter}{0}
\begin{figure}\centering%
\gdef\global@pandoccrossref@subfigures@footnotes{}%
\DeclareRobustCommand{\footnote}[1]{\footnotemark%
\stepcounter{pandoccrossref@subfigures@footnote@counter}%
\ifx\global@pandoccrossref@subfigures@footnotes\empty%
\gdef\global@pandoccrossref@subfigures@footnotes{{##1}}%
\else%
\g@addto@macro\global@pandoccrossref@subfigures@footnotes{, {##1}}%
\fi}}%
{\end{figure}%
\addtocounter{footnote}{-\value{pandoccrossref@subfigures@footnote@counter}}
\@for\f:=\global@pandoccrossref@subfigures@footnotes\do{\stepcounter{footnote}\footnotetext{\f}}%
\gdef\global@pandoccrossref@subfigures@footnotes{}}
\@ifpackageloaded{float}{}{\usepackage{float}}
\floatstyle{ruled}
\@ifundefined{c@chapter}{\newfloat{codelisting}{h}{lop}}{\newfloat{codelisting}{h}{lop}[chapter]}
\floatname{codelisting}{Listing}

\makeatother
\ifLuaTeX
  \usepackage{selnolig}  
\fi
\IfFileExists{bookmark.sty}{\usepackage{bookmark}}{\usepackage{hyperref}}
\IfFileExists{xurl.sty}{\usepackage{xurl}}{} 
\hypersetup{
  pdftitle={Quality Control of Image Sensors using Gaseous Tritium Light Sources},
  pdfauthor={David McFadden1,2,3; Brad Amos4; Rainer Heintzmann1,2,3},
  pdfkeywords={calibration, gain, conversion factor, quantum
efficiency, ccd, scmos},
  hidelinks,
  pdfcreator={LaTeX via pandoc}}

\title{Quality Control of Image Sensors using Gaseous Tritium Light
Sources}
\author{David McFadden\textsuperscript{1,2,3} \and Brad Amos\textsuperscript{4} \and Rainer Heintzmann\textsuperscript{1,2,3}}
\date{}

\begin{document}
\maketitle
\begin{abstract}
We propose a practical method for radiometrically calibrating cameras
using widely available gaseous tritium light sources
(\emph{betalights}). Along with the gain (conversion factor) and read
noise level, the predictable photon flux of the source allows us to
gauge the quantum efficiency. The design is easily reproducible with a
3D printer (three-dimensional printer) and three inexpensive parts.
Suitable for common image sensors, we believe that the method has the
potential to be a useful tool in microscopy facilities and optical labs
alike.
\end{abstract}

\textsuperscript{1} Institute of Physical Chemistry and Abbe Center of
Photonics, Friedrich-Schiller-University, Jena, Germany\\
\textsuperscript{2} Leibniz Institute of Photonic Technology,
Albert-Einstein-Straße 9, 07745 Jena, Germany\\
\textsuperscript{3} Jena Center for Soft Matter (JCSM), Friedrich
Schiller University Jena, Jena, Germany\\
\textsuperscript{4} Medical Research Council, MRC, Laboratory of
Molecular Biology, Cambridge, United Kingdom

\hypertarget{introduction}{%
\section{Introduction}\label{introduction}}

Photodetectors and cameras are common instruments used for acquiring
scientific data. Scientists are often particularly concerned about the
read noise performance and the photon conversion factor, which may vary
from device to device. Manufacturers therefore often include such
information in quality control reports provided to the customers upon
delivery. The detectors may, however, exhibit damage or ageing effects
which can affect the quality and reproducibility of data. Various robust
methods for end-user calibration exist {[}1{]}. But despite this, many
labs do not regularly perform quantitative quality control checks on
their instruments {[}2{]}. It is even less common to measure the
radiometric quantum efficiency in a laboratory setting. One explanation
could be a relative lack of convenient and low-cost calibration sources.

\hypertarget{existing-methods}{%
\section{Existing methods}\label{existing-methods}}

Measuring the gain and noise characteristics of a detector requires a
stable light source. In order to measure the quantum efficiency, the
light source must furthermore be calibrated to a radiometric standard.
This is the main application for calibrated radiometric transfer
standards {[}3{]}. In the visible spectrum, these are realised using
quartz-tungsten-halogen lamps, which have an irradiance uncertainty in
the range of 1\% {[}4{]}. They offer a limited lifetime, require a long
start-up period and a regular recalibration is also recommended {[}5{]}.
These lamps, in turn, require a calibrated power supply, resulting in a
combined setup that is both bulky and expensive, with prices on the
order of 1000 EUR to 6000 EUR. For most labs, a calibration tool that
costs as much as a camera itself is unlikely to be considered a good
return on investment.

Alternatively, one could perform a control experiment using a calibrated
detector and correct for the different detector geometries, but this
adds complexity to the procedure and introduces a potential for human
error. This is the method outlined in the industry-standard
EMVA1288{[}6{]}.

X-rays from radioactive sources, such as Fe55, are sometimes used to
measure the gain of visible range image sensors{[}7 ch.~2.3,8{]}. The
method can be challenging to use in situ, though, as the glass windows
of cooled cameras absorb X-rays. It is also not suited to measure the
visible range quantum efficiency as this requires visible light.

\hypertarget{gaseous-tritium-light-sources}{%
\section{Gaseous Tritium Light
Sources}\label{gaseous-tritium-light-sources}}

Gaseous Tritium Light Sources (GTLSs, or betalights), used in niche
products such as high-end watches, gun sights and fishing tackle, are
relatively inexpensive (approximate cost: 5-20 EUR) and emit a stable
and predictable light over long periods of time, ideally suited as a
standard light source. They rely on free electrons from the beta-decay
of tritium (hydrogen-3) gas to excite light-emitting phosphors that coat
the inside layer of a glass tube (figure \ref{fig:tube_photo}).

\begin{pandoccrossrefsubfigures}

\subfloat[]{\includegraphics[width=\textwidth,height=40mm]{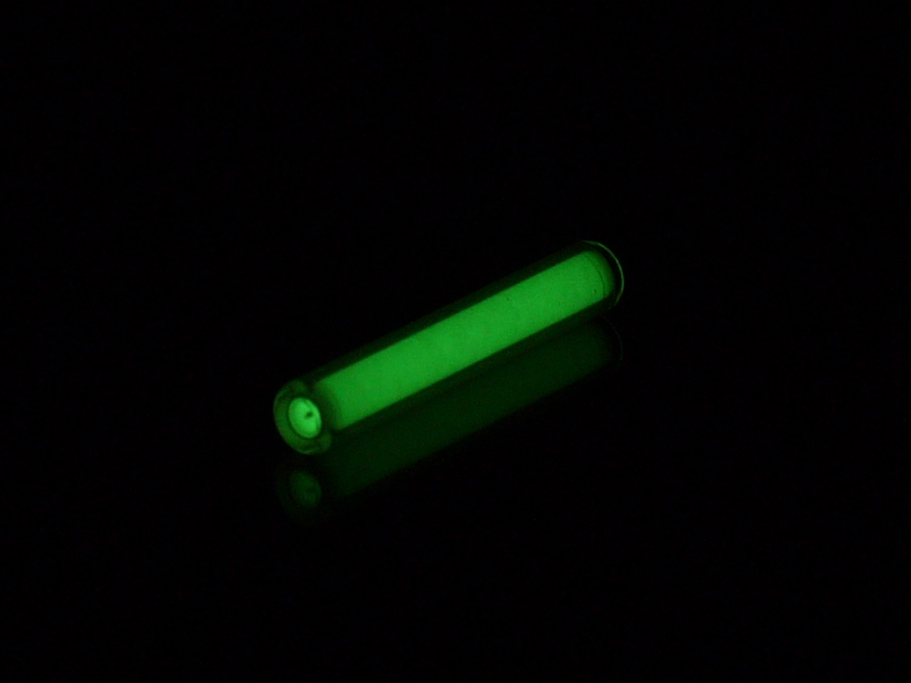}\label{fig:tube_photo}}
\subfloat[]{\includegraphics[width=\textwidth,height=40mm]{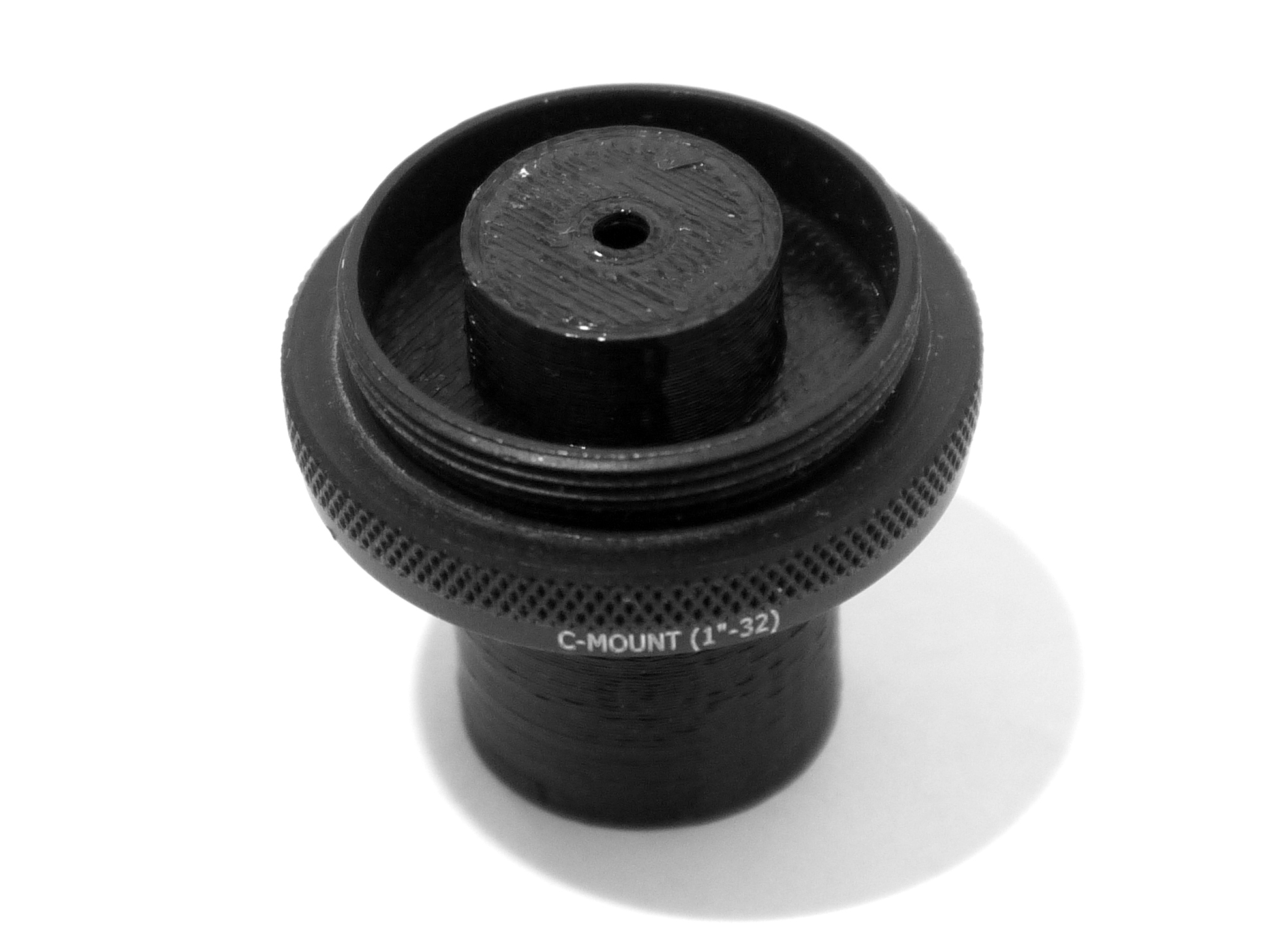}\label{fig:holder_photo}}

\caption[{Left: View of a 2 mm \(\times\) 12 mm tube. Right: Photograph
of the assembled source.}]{Left: View of a 2 mm \(\times\) 12 mm tube.
Right: Photograph of the assembled source.}

\label{fig:tube_frame}

\end{pandoccrossrefsubfigures}

Compared to thermal lamps, they offer some attractive properties: They
can be made very compact, being smaller than the hands on a watch. And
they are self-powered, independent of an electrical power supply.

The idea of using radioluminescent light sources as a low-light
radiometric standard is not new {[}9,10{]}. But while
radioluminescence-based sensor quality-control has been performed by
astronomical observatories {[}11{]}, radioluminescence based calibration
methods have not to our knowledge been widely adopted in microscopy
facilities.

While the use of radioactive materials initially raises safety concerns,
the quantities used in small GTLSs are considered safe and unlikely to
result in a significant radiation exposure, even in the worst-case
scenario {[}12,13{]}.

The regulatory situation is an important consideration for any
application. But while radionuclides, including tritium, are a
controlled substance across the world, many jurisdictions have
exemptions for devices up to a certain level of activity that would
usually include small GTLSs. For instance, in Germany, this limit is set
at one gigabecquerel {[}14{]}.

\hypertarget{stability}{%
\section{Stability}\label{stability}}

To assess the stability, we measured the signal that a green GTLS
produced on a calibrated photodiode (Thorlabs S130C sensor head with
PM100D power meter) at a fixed position over time. While work by
Mikhalchenko et al. {[}15{]} has demonstrated in principle the stability
of tritium radioluminescent sources, and others have praised their
stable properties {[}11{]}, {[}16{]} , we found little quantitative
data, and it is unclear how these results would apply to generically
sourced GTLSs, given their proprietary formulations and manufacturing
methods.

\begin{figure}
\hypertarget{fig:decay}{%
\centering
\includegraphics[width=89mm,height=\textheight]{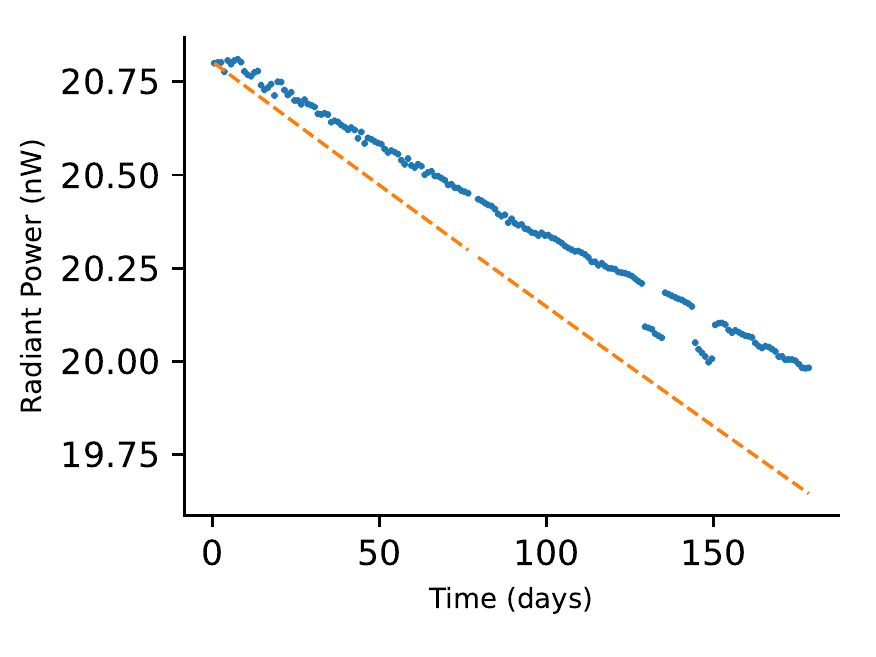}
\caption{A plot showing the continuous decrease in optical power over
time. Each dot represents one day's measurement. The sudden signal drop
by 0.6\% at 129 days and 144 days can likely be attributed to a
temperature decrease in the lab over the holiday period which can, in
turn, affect the photodiode. The root-mean-square deviation of a linear
fit is \(1.3\times10^{-3}\). If we exclude the aforementioned
temperature drop as outlier measurements, this decreases to
\(4.8\times10^{-4}\). The orange line represents the decay that we would
expect from the literature value for the radioactive decay of tritium
(half-life: 12.3 years).}\label{fig:decay}
}
\end{figure}

We purchased a green-emitting GTLS from a commercial vendor \footnote{Mad
  Nuclear Scientist Ltd., London, United Kingdom} and monitored the
optical power over 178 days (see section ``Optical Power measurement''
in the methods supplement). The device was found to have excellent
predictability. The decay is in good agreement with the expected decay
of tritium activity (half-life 12.3 years {[}17{]})(figure
\ref{fig:decay}). The output power decreased by 3.9\% over the duration
of the measurement, whereas the expected decay in tritium activity over
this time is 2.7\%. This corresponds to a half-life of 8.4 years. We
cannot rule out that this discrepancy is caused by a long-term drift in
the photodiode or the current meter. (cf.~``Power meter reliability'' in
the methods supplement).

Barring any nonlinear ageing process in the phosphor coating, these
results imply that the source could be used as a standard for low-light
optical power over many years.

It has been noted that the temperature dependence of GTLSs is a drawback
{[}18{]}, but we consider the mentioned \(0.3\% / ^{\circ} C\) to be
comparatively small for our intended application.

\hypertarget{design}{%
\section{Design}\label{design}}

With this knowledge, we conceptualised a radiometric calibration tool
for cameras. The goal is that it should be reliable, reproducible and
affordable. Using it to calibrate a camera should be a simple plug and
play procedure.

We designed a 3D (three-dimensional) printable mount (figure
\ref{fig:holder_photo}) for the aforementioned GTLS. Fused filament
fabrication 3D printers were used for development and manufacture
(Ultimaker 2/3, Prusa i3).

The design mounts the GTLS in a fixed position, and two circular
apertures define a divergent beam which exits from the detector-pointing
side (figure \ref{fig:source_detector}). The apertures are dimensioned
specifically to shape a beam that under-fills the detector's active area
at a chosen distance (figure \ref{fig:spot}). The extended size of the
first aperture results in a feathered edge for the beam spot on the
image sensor, yielding a continuous range of pixel intensities. This is
an ideal image pattern for the camera calibration method discussed in
the methods supplement (section ``Image analysis'').

Under-filling the detector also greatly simplifies the comparison of
quantum efficiencies between different detectors, as it is not necessary
to correct for different geometries if all of them are under-filled.

There are no optical elements between the source and the camera,
eliminating potential sources of misalignment and contamination. To
ensure a high-quality light seal and repeatable attachment, we fastened
it in a conventional c-mount adapter.

A correctly chosen geometry will also restrict the light's angle of
incidence on the detector, which is useful to ensure that the
measurements are representative of typical imaging applications.

For printing, we used both PLA (polylactic acid) and ABS (acrylonitrile
butadiene styrene), two common 3D printing materials (brand: BASF
Ultrafuse, black colour option). Both materials have a higher
transmission in the near-infrared than they do in the visible range
(cf.~methods supplement, section ``Light absorption of printed
materials''). This could bias measurements, as silicon-based detectors
are sensitive up to a wavelength of about 1200 nm. While we observed
that the ABS material had an excellent overall optical extinction, the
PLA material leaked a small amount of infrared light even through very
thick layers. Based on this, we would recommend that ABS be preferred
for printed parts. These absorption characteristics may vary from vendor
to vendor, however. Whilst PLA can still be used, measurements should be
performed away from NIR sources such as daylight and incandescent lamps,
and with the room lighting turned off as a precaution. This should be
enough to limit light leakage to very low levels. All printed parts
should be visually inspected for holes and defects.

The design is freely available at
https://github.com/mcfaddendavid/betalight-calibration{[}19{]}.
Parameters can be tuned so that the spot best matches various detector
geometries. The bill of materials for non-printed parts consists of:

\begin{longtable}[]{@{}
  >{\raggedright\arraybackslash}p{(\columnwidth - 4\tabcolsep) * \real{0.0870}}
  >{\raggedright\arraybackslash}p{(\columnwidth - 4\tabcolsep) * \real{0.7500}}
  >{\raggedright\arraybackslash}p{(\columnwidth - 4\tabcolsep) * \real{0.1630}}@{}}
\toprule()
\begin{minipage}[b]{\linewidth}\raggedright
Number
\end{minipage} & \begin{minipage}[b]{\linewidth}\raggedright
Description
\end{minipage} & \begin{minipage}[b]{\linewidth}\raggedright
Approx. Price
\end{minipage} \\
\midrule()
\endhead
1 & Thorlabs C-mount retaining ring, Part number: CMRR & 11.52 EUR \\
1 & Thorlabs CS- to C-Mount Extension Adapter, Part number: CML05 &
16.78 EUR \\
1 & Cylindrical GTLS light source 2 mm x 12 mm (green colour suggested)
& 8.80 EUR \\
\bottomrule()
\end{longtable}

\begin{pandoccrossrefsubfigures}

\subfloat[]{\includegraphics[width=\textwidth,height=40mm]{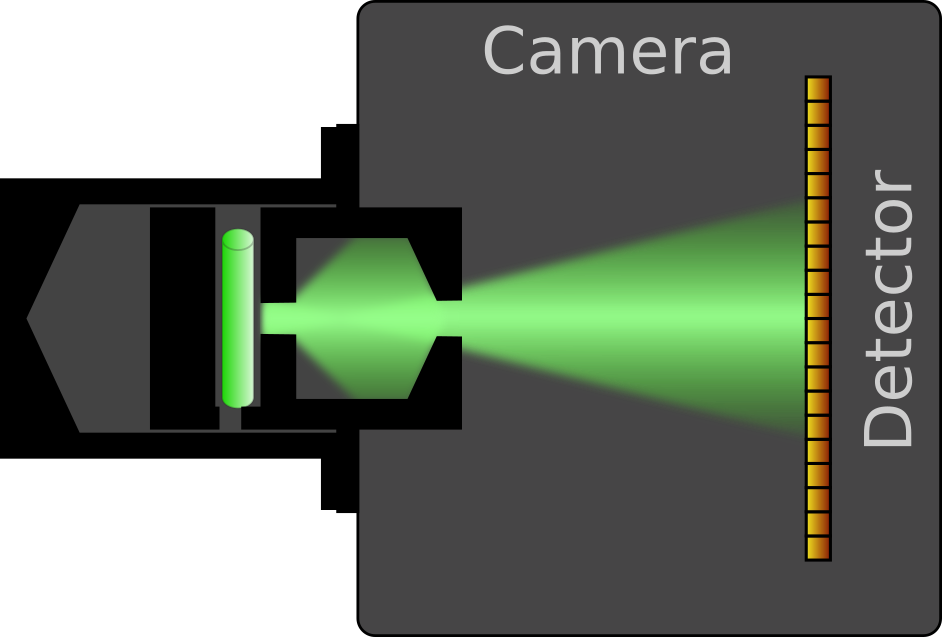}\label{fig:source_detector}}
\subfloat[]{\includegraphics[width=\textwidth,height=40mm]{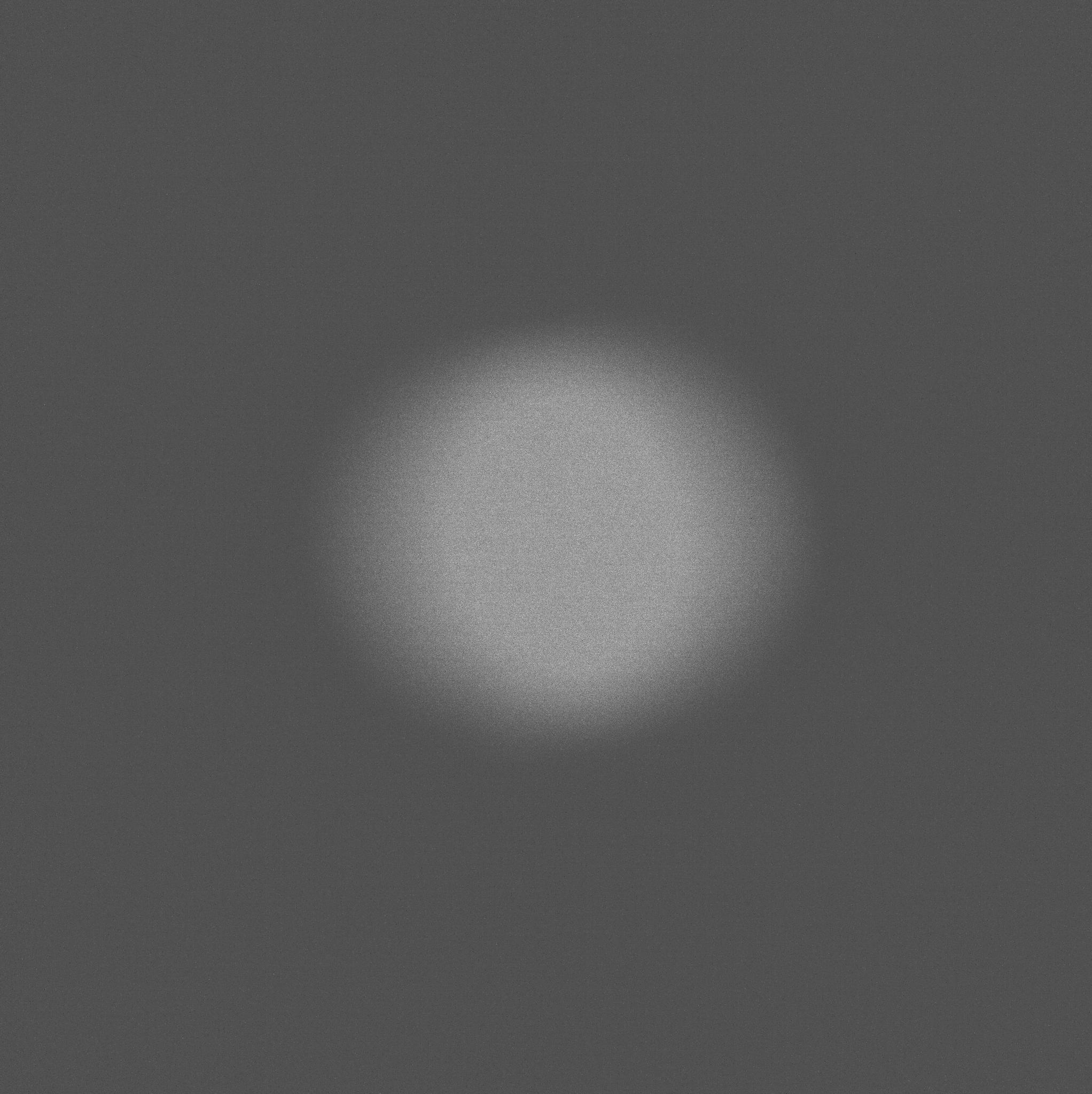}\label{fig:spot}}

\caption[{Left: Cross-section showing the mounted light source attached
to a camera. The divergent light beam (NA approximately 0.1) impinges
upon the image sensor. Apertures between the GTLS tube and the detector
plane restrict the beam size to under-fill the detector surface. Right:
The beam profile on the image sensor (physical size 16.4 mm \(\times\)
16.4 mm). Essentially all of the light is confined within the sensor
area whilst covering a good portion of the available space.}]{Left:
Cross-section showing the mounted light source attached to a camera. The
divergent light beam (NA approximately 0.1) impinges upon the image
sensor. Apertures between the GTLS tube and the detector plane restrict
the beam size to under-fill the detector surface. Right: The beam
profile on the image sensor (physical size 16.4 mm \(\times\) 16.4 mm).
Essentially all of the light is confined within the sensor area whilst
covering a good portion of the available space.}

\label{fig:crossec_frame}

\end{pandoccrossrefsubfigures}

\hypertarget{camera-calibration-and-charge-carriers}{%
\section{Camera calibration and charge
carriers}\label{camera-calibration-and-charge-carriers}}

First, the mounted source is screwed onto the camera lens mount, and a
stack of images at constant exposure is acquired (cf. figure
\ref{fig:spot}). Afterwards, the source is replaced with a camera body
cap and we acquire a stack of dark exposures with identical camera
settings.

\begin{figure}
\hypertarget{fig:ptc}{%
\centering
\includegraphics[width=130mm,height=\textheight]{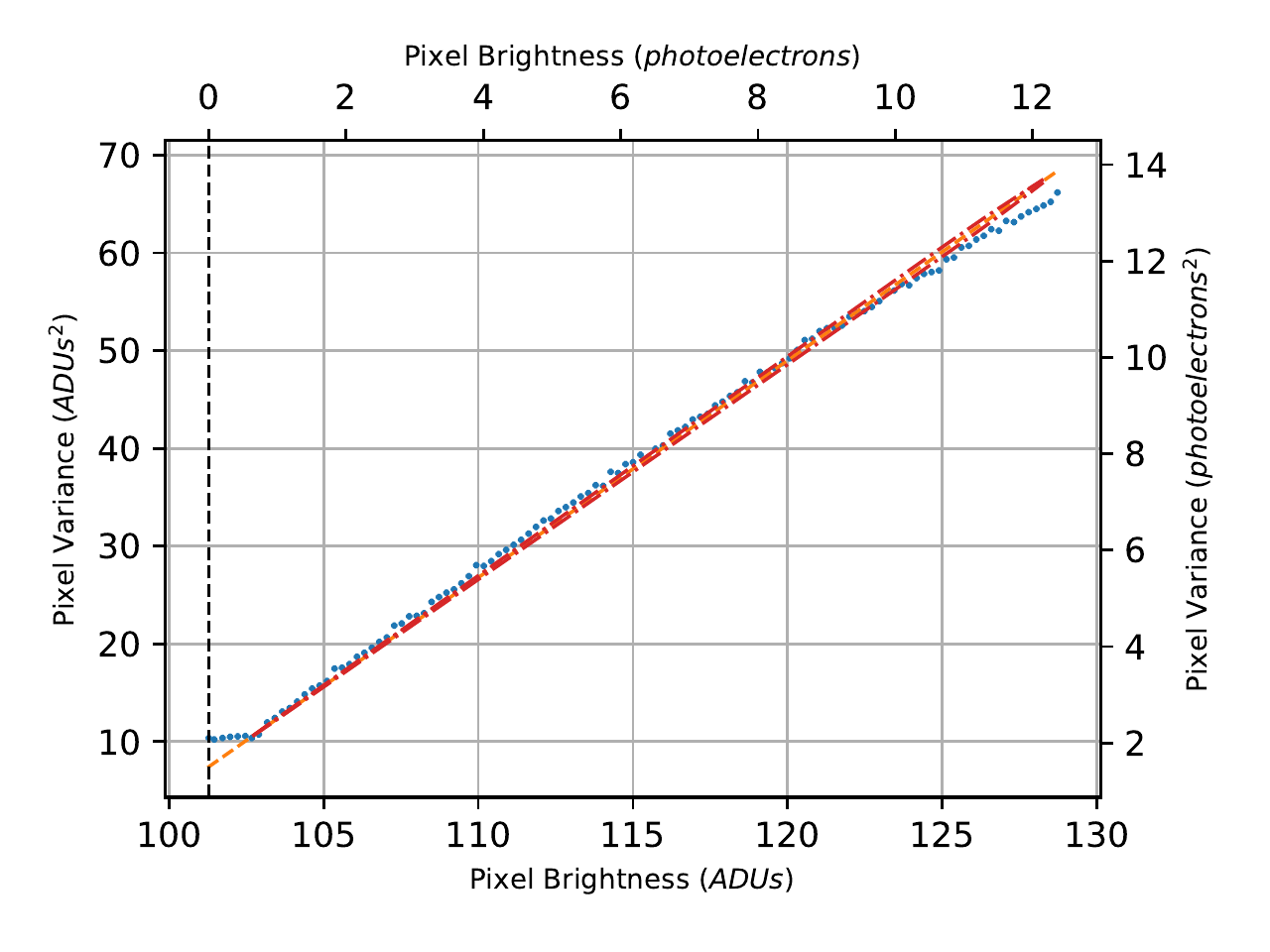}
\caption{A photon transfer curve for an sCMOS camera generated by the
described method using the GTLS calibration source (exposure time 20
ms). The blue dots represent the pixel bins, and the fitted line is in
orange. Surrounding the fit are lines representing the standard
deviation of the variance estimates. The gain estimate is 0.45 electrons
per digital unit. The read noise is 1.69 electrons
root-mean-square.}\label{fig:ptc}
}
\end{figure}

The typical approach for calibrating gain is the photon transfer method
{[}1,20,21{]}. Our implementation is based on a mean-variance curve
(figure \ref{fig:ptc}) and is described in the electronic supplementary
material. We use it to obtain estimates for the read noise, the offset,
and the camera gain, which relates the arbitrary analog-digital-units
(ADU) in the output image to a specific number of charge carriers
(photoelectrons) and is expressed in electrons per ADU. Expressing the
noise in terms of effective charge carriers is the only meaningful way
to compare cameras with different designs.

Subtracting the mean projection along the time axis of the dark exposure
stack from the mean projection of the bright exposure gives us an
unbiased average frame. We sum the signal \(S(i)\) of all pixels \(i\)
and multiply the result with the camera gain. This yields a value for
the average number of detected photoelectrons per exposure \(N_{e^-}\)

\[ N_{e^-} = \sum_{i} S(i) \times \textrm{gain} \]

Dividing this by the exposure time gives us a rate of effective
photoelectrons \(\Phi(e^-)\):

\begin{equation}\protect\hypertarget{eq:electron_flux}{}{\Phi(e^-) = \frac{N_{e^-}}{\textrm{exposure time}}}\label{eq:electron_flux}\end{equation}

\hypertarget{sec:spectro-pm}{%
\section{Source characterisation}\label{sec:spectro-pm}}

\begin{figure}
\hypertarget{fig:spec}{%
\centering
\includegraphics[width=89mm,height=\textheight]{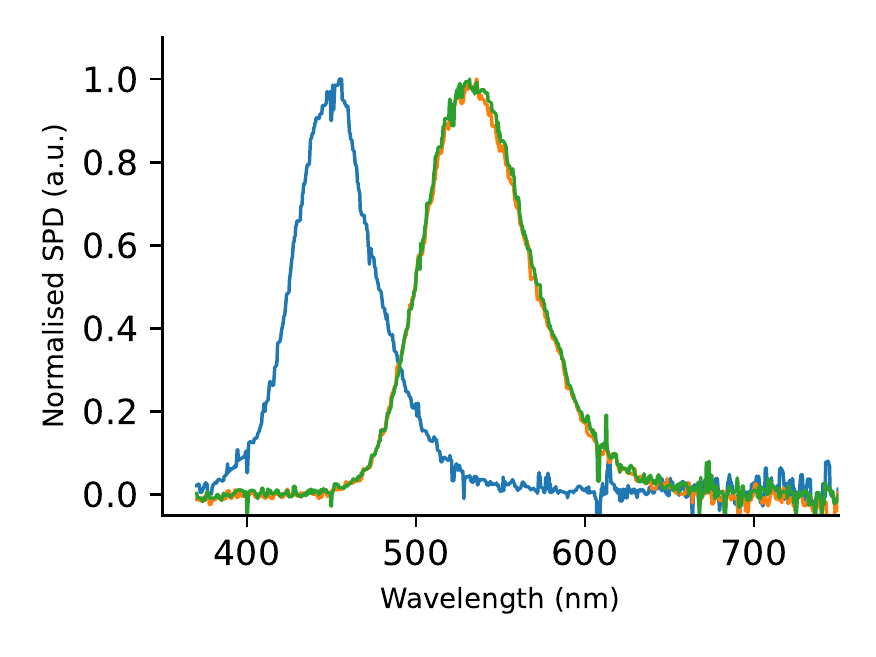}
\caption{Normalised power density spectra obtained from three different
GTLSs, two green (green and orange curves) and one blue (blue curve),
with power weighted mean wavelengths at 537 nm, 541 nm and 461 nm
respectively. The green spectra closely resemble each other with the
results differing by only 0.6\%. Repeated measurements on the same
source achieved a standard deviation of 0.7\%. Additionally, we measured
the green source after four months and could not detect a significant
change in the result, the measurements differing by 0.8\%. (Integration
time 12 seconds, averaging over ca. 514 acquired spectra. Analysed
range: 370nm to 750nm)}\label{fig:spec}
}
\end{figure}

A radiometrically calibrated spectrometer yields the spectral power
distribution of the source. The diffuse emission and low light level are
an additional challenge for some low-sensitivity uncooled spectrometers.
Fibre-based spectrometers may benefit from using a fibre-based mount for
the GTLS (figure \ref{fig:fibre}). Using careful background subtraction,
we were nevertheless able to obtain a satisfactory spectrum with a
compact uncooled spectrometer (figure \ref{fig:spectrometer}) using only
free-space propagation (figure \ref{fig:spec}).

\begin{pandoccrossrefsubfigures}

\subfloat[]{\includegraphics[width=60mm,height=\textheight]{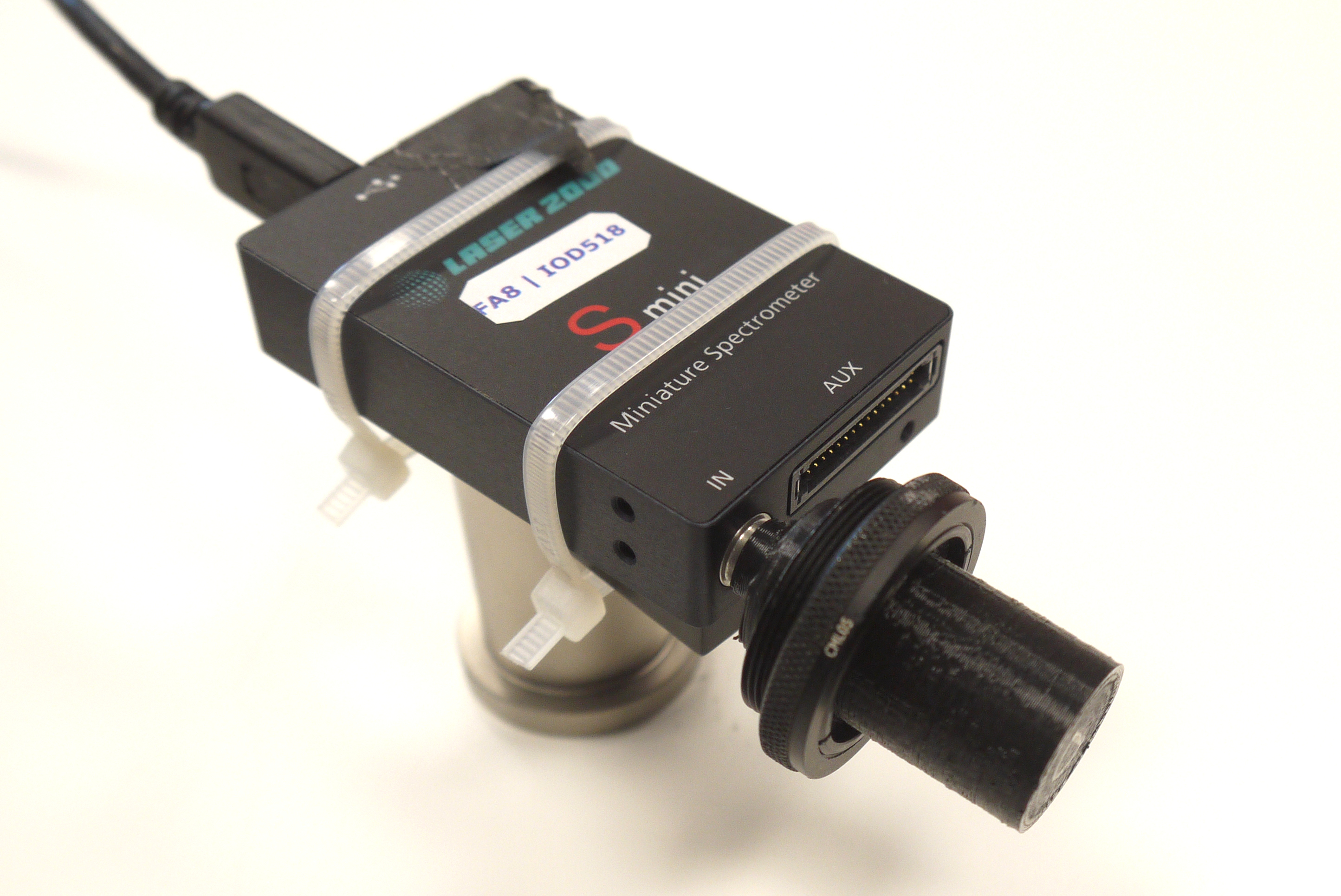}\label{fig:spectrometer}}
\subfloat[]{\includegraphics[width=40mm,height=\textheight]{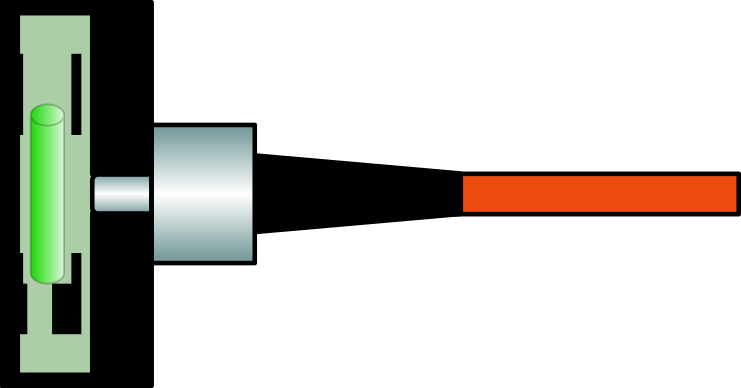}\label{fig:fibre}}

\caption[{Left: The source is attached as close as possible to the input
of a spectrometer (Qmini, RGB Photonics GmbH). Right: A mount designed
to couple light from the GTLS to an optical fibre. This usually improves
the signal level for fibre-based spectrometers.}]{Left: The source is
attached as close as possible to the input of a spectrometer (Qmini, RGB
Photonics GmbH). Right: A mount designed to couple light from the GTLS
to an optical fibre. This usually improves the signal level for
fibre-based spectrometers.}

\label{fig:figureRef}

\end{pandoccrossrefsubfigures}

A calibrated spectrometer will output a spectral power distribution
\(\textrm{SPD}(\lambda)\). In SI units, this will be in W/nm. Using this
as a weighting function allows us to characterise the source by a
power-weighted mean wavelength \(\bar{\lambda}\):

\[ \bar{\lambda} = \frac{\int_{\lambda} \lambda \: \textrm{SPD}(\lambda) d \lambda}{\int_{\lambda} \textrm{SPD}(\lambda) d \lambda} \]

The assembled calibration source is then mounted pointing towards a
power-calibrated photodiode. The distance to the photodiode surface is
exactly the same that the source would otherwise be from the surface of
the image sensor of a c-mount camera body. (see methods supplement).

We measure the photocurrent that the beam induces on the under-filled
detector, and by correcting for the spectral responsivity of the
photodiode as well as the spectral power distribution of the source
\(\textrm{SPD}(\lambda)\) (see methods supplement, section ``Optical
Power measurement''), we obtain the expected optical power \(P\) of the
beam.

The power-weighted mean wavelength can be associated with a photon
energy:

\[E_{ph} = \frac{hc}{\bar{\lambda}}\]

Dividing the power \(P\) by this photon energy yields the expected rate
of impinging photons \(\Phi(\textrm{ph})\):

\begin{equation}\protect\hypertarget{eq:photon_flux}{}{ \Phi(\textrm{ph}) = \frac{P}{E_{\textrm{ph}}} }\label{eq:photon_flux}\end{equation}

Note that using the photon energy of the power-weighted mean wavelength
to calculate the photon rate in this way is valid for any temporally
stable spectral shape. If, however, the spectral shape experiences a
shift over time, then it will result in a different photon rate. Such a
change might occur in GTLSs with multi-coloured or blended phosphors.

We found that our photodiode power meter delivered satisfactory and
repeatable results, with a total optical power \(E_{ph}\) of 118 pW.
With a \(\bar{\lambda}\) of 537 nm, this corresponds to a photon flux
\(\Phi(\textrm{ph})\) of \(3.20\times10^{8} s^{-1}\).

The quantum efficiency is simply the ratio between the rate of detected
photoelectrons \(\Phi(e^-)\) (eq.~\ref{eq:electron_flux}) and the photon
flux \(\Phi(\textrm{ph})\)(eq.~\ref{eq:photon_flux}):

\[ \textrm{Quantum efficiency} = \frac{\Phi(e^-)}{\Phi(ph)} \]

\hypertarget{results}{%
\section{Results}\label{results}}

Testing the procedure on three scientific cameras, we obtain quantum
efficiency values that agree very well with the quantum efficiencies
quoted by the manufacturers\footnote{The data was read manually from
  graphs provided by the manufacturers. The manufacturers do not provide
  error estimates. We assume a 3\% relative uncertainty in the
  calibrations as well as a 2\% absolute uncertainty in the rendering of
  the quantum efficiency graphs. There are more potential unknown errors
  which we cannot account for (see methods supplement, section
  ``Manufacturer stated quantum efficiency''). For its working range,
  the manufacturer quotes the uncertainty of the power meter as ±3\%
  above 450nm, and ±5\% below 450nm {[}22{]}. However, it should be
  noted that we are operating the photodiode below its specified power
  range (Our source has an optical power of 118 pW, whereas the
  specified minimum power of the photodiode is 500pW, cf.~methods
  supplement, section ``Power meter reliability''). The spectral
  distribution of the source is another crucial input, and measurements
  on the green source suggest an uncertainty on the order of 1\%. We add
  this to a 3\% photodiode uncertainty to obtain a relative error of 4\%
  for our measurements.}, once the spectral power distribution of the
sources is taken into account. Two of these are based on scientific CMOS
(sCMOS) sensors. CMOS sensors are active pixel sensors named after the
complementary metaloxide-semiconductor (CMOS) fabrication method, and
the ``scientific'' is commonly used to distinguish high-performance
scientific cameras from low-performance sensors common in consumer
devices. The third sensor is an electronmultiplying charge coupled
device (EMCCD) sensor operated in its non-amplified mode. Table
\ref{tbl:qe_tab} shows values obtained using both blue and green
coloured GTLSs. We find that all but one of the measurements (denoted
with ``*``) fall within the estimated uncertainty intervals of the
manufacturer-specified values.

\hypertarget{tbl:qe_tab}{}
\begin{longtable}[]{@{}lllll@{}}
\caption{\label{tbl:qe_tab}Quantum efficiency values obtained from the
calibration routine for three cameras along with values quoted by their
manufacturers. All but one of the measurements (denoted with ``*``) fall
within the estimated uncertainty intervals of the manufacturer-specified
values. The central wavelengths of the blue and green sources were at
461 nm and 537 nm respectively.}\tabularnewline
\toprule()
& Colour & Measured & Quoted & Difference \\
\midrule()
\endfirsthead
\toprule()
& Colour & Measured & Quoted & Difference \\
\midrule()
\endhead
EMCCD camera & Green & 88.6\%±3.5\% & 89.3\%±4.7\% & -0.7\% \\
& Blue & 73.8\%±3.0\% & 74.1\%±4.3\% & +1.2\% \\
sCMOS Camera no. 1 & Green & 69.0\%±2.8\% & 70.8\%±4.2\% & -1.8\% \\
& Blue & 53.7\%±2.1\% & 55.5\%±3.7\% & -1.8\% \\
sCMOS Camera no. 2 & Green & 70.5\%±2.8\% & 78.5\%±4.4\% & -8.0\% (*) \\
& Blue & 66.8\%±2.7\% & 73.0\%±4.2\% & -6.2\% \\
\bottomrule()
\end{longtable}

Ultimately the results will also hinge upon the validity of the photon
transfer method. It is therefore paramount to scrutinise the results of
a photon transfer curve before using them in further processing. Some of
our scientific cameras produced repeatable curves across a wide dynamic
range, and we would have more confidence in such results. Figure
\ref{fig:ptc} is such a curve. Other cameras produced nonlinear curves
in a way that would clearly invalidate gain calibration results. (see
methods supplement).

\hypertarget{conclusion}{%
\section{Conclusion}\label{conclusion}}

We have demonstrated an affordable and suitably accurate tool for
calibrating cameras in a plug-and-play fashion in situ. The design
requires only minimal parts and a 3D printer, which allows it to be
quickly disseminated to labs where it might be useful.

Absolute quantum efficiency measurement was also demonstrated, but this
requires additional spectral information obtained from a radiometrically
calibrated spectrometer. Note that a radiometrically accurate spectrum
is generally required for conventional quantum-efficiency measurement
methods, and the EMVA1288 standard specifically recommends verifying the
source spectrum using a (radiometrically calibrated) spectrometer
{[}6{]}. The standard does, however, allow users to forgo this
verification if an accurate spectrum can be provided by the light source
manufacturer.

We found no significant difference in the spectrum between two different
green GTLSs and were able to reproduce the same spectrum four months
later. This suggests that sufficiently accurate prior knowledge of the
source spectrum could be provided. But as our GTLSs are unbranded, we
cannot know what quality control mechanisms they underwent and the
vendor also supplied very limited documentation. As such, we can't claim
to have confidence that our spectral data will apply to GTLSs purchased
by others. We would therefore recommend that anybody wishing to
reproduce the method measure the spectrum of the GTLS upon delivery and
assembly, and periodically thereafter in order to monitor for changes.

Compared to thermal calibration lamps, the optical power of GTLSs is
very weak, which prevents us from measuring across the full dynamic
range of most non-amplified detectors. Nevertheless, it is likely to be
the appropriate power range for many life-science applications,
especially single-molecule-localisation microscopy.

The relatively broad spectrum (FWHM \(\approx\) 100 nm) of the source
means that we are averaging over a wider spectral range than the 50 nm
permitted by EMVA1288 {[}6{]}. Conversely, for some applications, the
spectral width might be considered too narrow or not at the correct
location. The availability of GTLSs with phosphors in many different
colours {[}23{]} suggests, however, that the method could be applied to
other spectral ranges. These sources would need to be evaluated
individually for their stability, and such an approach might not have
any inherent advantage over existing tools for relative polychromatic
calibration, such as described by JM Beach {[}24{]}.

As we do not expose the entire image sensor, local defects at the edges
go undetected, though these can also be easily found with conventional
light sources or by using a mount without baffles for the GTLS.

It should be noted that our method is especially useful for assessing
the long-term stability of detectors. The calibration source should be
stored in a dust-free container to ensure that this is not thwarted by
dirt or dust blocking the light path.

The current design also requires the camera to be detached from any lens
mount, which is a drawback as many microscope users do not wish to
remove the cameras from their setups once they have been assembled.

We are currently working on a design that could alleviate this by
mounting the source in an alternative location in the optical path of an
instrument. For microscopes, this could be at the sample plane or the
microscope objective mount. This would allow us to assess the long term
stability of the entire optical path of the instrument.

\hypertarget{supplementary-information}{%
\paragraph{Supplementary information}\label{supplementary-information}}

Electronic supplementary material is available online at
https://doi.org/10.6084/m9.figshare.c.5768142

\hypertarget{data-accessibility}{%
\paragraph{Data accessibility}\label{data-accessibility}}

The data, as well as the analysis script that was used to produce the
results of this article, are available on the Dryad data repository with
the following DOI: https://doi.org/10.5061/dryad.cvdncjt5f The code for
the image analysis, as described in the methods supplement, is
maintained in the freely available ``NanoImagingPack'' library:
https://gitlab.com/bionanoimaging/nanoimagingpack/ 3D files for the
printed parts of the calibration source, as well as assembly
instructions, are available at:
https://github.com/mcfaddendavid/betalight-calibration

\hypertarget{contributions}{%
\paragraph{Contributions}\label{contributions}}

David McFadden designed, coordinated and performed the experiments,
developed the designs, analysed the results and drafted the manuscript.
Brad Amos came up with the idea of using betalights for camera
calibration and revised the manuscript. Rainer Heintzmann participated
in the analysis and revised the manuscript. All authors gave final
approval for publication and agree to be held accountable for the work
performed therein.

\hypertarget{competing-interests}{%
\paragraph{Competing interests}\label{competing-interests}}

We declare we have no competing interests.

\hypertarget{funding}{%
\paragraph{Funding}\label{funding}}

This work was supported by the Deutsche Forschungsgemeinschaft (1278
Polytarget, Project C04).

\hypertarget{references}{%
\section*{References}\label{references}}
\addcontentsline{toc}{section}{References}

\hypertarget{refs}{}
\begin{CSLReferences}{0}{0}
\leavevmode\vadjust pre{\hypertarget{ref-heintzmannCalibratingPhotonCounts2018}{}}%
\CSLLeftMargin{1. }%
\CSLRightInline{Heintzmann R, Relich PK, Nieuwenhuizen RPJ, Lidke KA,
Rieger B. 2018 \href{http://arxiv.org/abs/1611.05654}{Calibrating photon
counts from a single image}. arXiv:161105654 [astro-ph, physics:physics] [Internet]}

\leavevmode\vadjust pre{\hypertarget{ref-boehmQUAREPLiMiCommunityEndeavor2021a}{}}%
\CSLLeftMargin{2. }%
\CSLRightInline{Boehm U \emph{et al.} 2021 {QUAREP-LiMi}: A community
endeavor to advance quality assessment and reproducibility in light
microscopy. \emph{Nature Methods} \textbf{18}, 1423--1426.
(doi:\href{https://doi.org/10.1038/s41592-021-01162-y}{10.1038/s41592-021-01162-y})}

\leavevmode\vadjust pre{\hypertarget{ref-hollandtPrimarySourcesUse2005b}{}}%
\CSLLeftMargin{3. }%
\CSLRightInline{Hollandt J, Seidel J, Klein R, Ulm G, Migdall A, Ware M.
2005 Primary sources for use in radiometry. In \emph{Experimental
{Methods} in the {Physical Sciences}} (eds AC Parr, RU Datla, JL
Gardner), pp. 213--290. {Amsterdam, The Netherlands}: {Elsevier}.
(doi:\href{https://doi.org/10.1016/S1079-4042(05)41005-X}{10.1016/S1079-4042(05)41005-X})}

\leavevmode\vadjust pre{\hypertarget{ref-ojanenDoublecoiledTungstenFilament2012}{}}%
\CSLLeftMargin{4. }%
\CSLRightInline{Ojanen M, Kärhä P, Nevas S, Sperling A, Mäntynen H,
Ikonen E. 2012 Double-coiled tungsten filament lamps as absolute
spectral irradiance reference sources. \emph{Metrologia} \textbf{49},
S53--S58.
(doi:\href{https://doi.org/10.1088/0026-1394/49/2/S53}{10.1088/0026-1394/49/2/S53})}

\leavevmode\vadjust pre{\hypertarget{ref-johnsonValidationDisseminationSpectral2012}{}}%
\CSLLeftMargin{5. }%
\CSLRightInline{Johnson BC, Graham GD, Saunders RD, Yoon HW, Shirley EL.
2012 Validation of the dissemination of spectral irradiance values using
{FEL} lamps. In (eds JJ Butler, X Xiong, X Gu), p. 85100E. {San Diego,
CA, USA}.
(doi:\href{https://doi.org/10.1117/12.930801}{10.1117/12.930801})}

\leavevmode\vadjust pre{\hypertarget{ref-EMVAStandard12882021}{}}%
\CSLLeftMargin{6. }%
\CSLRightInline{European Machine Vision Association. 2021 \emph{{EMVA
Standard} 1288 {Standard} for {Characterization} of {Image Sensors} and
{Cameras Release} 4.0 {General}}. {Spain}: {European Machine Vision
Association}. See
\url{https://www.emva.org/standards-technology/emva-1288/}.}

\leavevmode\vadjust pre{\hypertarget{ref-janesickScientificChargeCoupledDevices2001}{}}%
\CSLLeftMargin{7. }%
\CSLRightInline{Janesick JR. 2001 \emph{Scientific {Charge-Coupled
Devices}}. {Bellingham, WA, USA}: {SPIE}.
(doi:\href{https://doi.org/10.1117/3.374903}{10.1117/3.374903})}

\leavevmode\vadjust pre{\hypertarget{ref-weatherillElectroopticalTestSystem2017}{}}%
\CSLLeftMargin{8. }%
\CSLRightInline{Weatherill DP, Arndt K, Plackett R, Shipsey IPJ. 2017 An
electro-optical test system for optimising operating conditions of {CCD}
sensors for {LSST}. \emph{J. Inst.} \textbf{12}, C12019--C12019.
(doi:\href{https://doi.org/10.1088/1748-0221/12/12/C12019}{10.1088/1748-0221/12/12/C12019})}

\leavevmode\vadjust pre{\hypertarget{ref-hanleRadiolumineszenzAlsLichtquelle1956}{}}%
\CSLLeftMargin{9. }%
\CSLRightInline{Hanle W, Kügler I. 1956 Radiolumineszenz als
{Lichtquelle Konstanter Intensität}. \emph{Optica Acta: International
Journal of Optics}
(doi:\href{https://doi.org/10.1080/713823667}{10.1080/713823667})}

\leavevmode\vadjust pre{\hypertarget{ref-yamamotoStandardLightSource1975a}{}}%
\CSLLeftMargin{10. }%
\CSLRightInline{Yamamoto O, JA, Takenaga M, JA, Tsujimoto Y, JA. 1975
\href{https://patft.uspto.gov/netacgi/nph-Parser?Sect1=PTO1\&Sect2=HITOFF\&d=PALL\&p=1\&u=\%2Fnetahtml\%2FPTO\%2Fsrchnum.htm\&r=1\&f=G\&l=50\&s1=3889124.PN.\&OS=PN/3889124\&RS=PN/3889124}{Standard
light source utilizing spontaneous radiation -{United States Patent}:
3889124}. }

\leavevmode\vadjust pre{\hypertarget{ref-amicoDetectorMonitoringProject2008}{}}%
\CSLLeftMargin{11. }%
\CSLRightInline{Amico P, Ballester P, Hummel W, LoCurto G, Lundin L,
Modigliani A, Sinclaire P, Vanzi L. 2008 The {Detector Monitoring
Project}. In \emph{The 2007 {ESO Instrument Calibration Workshop}} (eds
A Kaufer, F Kerber), pp. 11--21. {Berlin, Germany}: {Springer}.
(doi:\href{https://doi.org/10.1007/978-3-540-76963-7_2}{10.1007/978-3-540-76963-7\_2})}

\leavevmode\vadjust pre{\hypertarget{ref-BfSRadioactiveMaterials2020}{}}%
\CSLLeftMargin{12. }%
\CSLRightInline{Bundesamt für Strahlenschutz. 2020 {BfS} - {Radioactive}
materials in watches. See
\url{https://www.bfs.de/EN/topics/ion/daily-life/watches/watches_node.html} . Archived at
\url{https://web.archive.org/web/20200608172547/https://www.bfs.de/EN/topics/ion/daily-life/watches/watches_node.html}
}

\leavevmode\vadjust pre{\hypertarget{ref-DecisionsAdoptionOECD1973}{}}%
\CSLLeftMargin{13. }%
\CSLRightInline{OECD Council. 1973 \emph{Decisions on the {Adoption} of
{OECD Legal Instruments Radiation Protection Standards} for {Gaseous
Tritium Light Devices}}. {Paris}: {Organisation for Economic
Co-operation and Development}. See
\url{https://legalinstruments.oecd.org/en/instruments/OECD-LEGAL-0110}.}

\leavevmode\vadjust pre{\hypertarget{ref-Strahlenschutzverordnung}{}}%
\CSLLeftMargin{14. }%
\CSLRightInline{Federal Republic of Germany. In press.
Strahlenschutzverordnung. See
\url{https://www.gesetze-im-internet.de/strlschv_2018/}
Archived at
\url{http://web.archive.org/web/20201002092905/https://www.gesetze-im-internet.de/strlschv_2018/
}
}

\leavevmode\vadjust pre{\hypertarget{ref-mikhalchenkoPropertiesRadioluminescenceSources2011}{}}%
\CSLLeftMargin{15. }%
\CSLRightInline{Mikhal'chenko AG. 2011 Properties of radioluminescence
sources for photometry. \emph{J. Opt. Technol.} \textbf{78}, 452.
(doi:\href{https://doi.org/10.1364/JOT.78.000452}{10.1364/JOT.78.000452})}

\leavevmode\vadjust pre{\hypertarget{ref-maretteRocketborneBaffledPhotometer1976}{}}%
\CSLLeftMargin{16. }%
\CSLRightInline{Marette G, Gérard J-C. 1976 Rocket-borne baffled
photometer: Design and calibration. \emph{Appl. Opt.} \textbf{15}, 437.
(doi:\href{https://doi.org/10.1364/AO.15.000437}{10.1364/AO.15.000437})}

\leavevmode\vadjust pre{\hypertarget{ref-lucasComprehensiveReviewCritical2000}{}}%
\CSLLeftMargin{17. }%
\CSLRightInline{Lucas LL, Unterweger MP. 2000 Comprehensive {Review} and
{Critical Evaluation} of the {Half-Life} of {Tritium}. \emph{J Res Natl
Inst Stand Technol} \textbf{105}, 541--549.
(doi:\href{https://doi.org/10.6028/jres.105.043}{10.6028/jres.105.043})}

\leavevmode\vadjust pre{\hypertarget{ref-abbottCCDsESOSystematic1995}{}}%
\CSLLeftMargin{18. }%
\CSLRightInline{Abbott TMC. 1995 The {CCDs} at {ESO}: {A Systematic
Testing Program}. In \emph{New {Developments} in {Array Technology} and
{Applications}} (eds AGD Philip, KA Janes, AR Upgren), pp. 343--344.
{Dordrecht, Netherlands}: {Springer Netherlands}.
(doi:\href{https://doi.org/10.1007/978-94-011-0383-1_52}{10.1007/978-94-011-0383-1\_52})}

\leavevmode\vadjust pre{\hypertarget{ref-mcfaddenGTLSbasedCalibrationSource2021}{}}%
\CSLLeftMargin{19. }%
\CSLRightInline{McFadden D. 2021 \emph{{GTLS-based} calibration source
for cameras}. See
\url{https://github.com/mcfaddendavid/betalight-calibration}.}

\leavevmode\vadjust pre{\hypertarget{ref-janesickPhotonTransfer2007}{}}%
\CSLLeftMargin{20. }%
\CSLRightInline{Janesick JR. 2007 \emph{Photon {Transfer}}. {1000 20th
Street, Bellingham, WA 98227-0010 USA}: {SPIE}.
(doi:\href{https://doi.org/10.1117/3.725073}{10.1117/3.725073})}

\leavevmode\vadjust pre{\hypertarget{ref-van1998digital}{}}%
\CSLLeftMargin{21. }%
\CSLRightInline{van Vliet L, Sudar D, Young I. 1998 Digital fluorescence
imaging using cooled charge-coupled device array cameras. In \emph{Cell
biology, second edition, volume {III}} (ed K Simons), pp. 109--120. {New
York, N.Y}: {Academic Press}. }

\leavevmode\vadjust pre{\hypertarget{ref-SpecSheetS130C}{}}%
\CSLLeftMargin{22. }%
\CSLRightInline{Thorlabs Inc. In press. Spec sheet {S130C}. See
\url{https://www.thorlabs.de/drawings/f7a637e286876c21-ACC02A8B-ED79-777B-5D7939344598B79F/S130C-SpecSheet.pdf}. Archived at 
\url{http://web.archive.org/web/20220623155622/https://www.thorlabs.de/drawings/f7a637e286876c21-ACC02A8B-ED79-777B-5D7939344598B79F/S130C-SpecSheet.pdf}.}

\leavevmode\vadjust pre{\hypertarget{ref-Trigalight2020}{}}%
\CSLLeftMargin{23. }%
\CSLRightInline{MB Microtec AG. 2020 Trigalight. See
\url{http://www.trigalight.com/} Archived at
\url{https://web.archive.org/web/20200603160423/http://www.trigalight.com/}.}

\leavevmode\vadjust pre{\hypertarget{ref-beachLEDLightCalibration1997}{}}%
\CSLLeftMargin{24. }%
\CSLRightInline{Beach JM. 1997 A {LED} light calibration source for
dual-wavelength microscopy. \emph{Cell Calcium} \textbf{21}, 63--68.
(doi:\href{https://doi.org/10.1016/S0143-4160(97)90097-X}{10.1016/S0143-4160(97)90097-X})}

\end{CSLReferences}

\end{document}